\documentstyle[multicol,aps]{revtex}

\def\simge{\,{}^>_{\sim}\,}
\def\simle{\,{}^<_{\sim}\,}

\begin{document}
\draft
\title{Could MACHOS be Primordial Black Holes formed during the QCD Epoch?}
\author{Karsten Jedamzik}
\address{Max-Planck-Institut f\"ur Astrophysik, 85740 Garching bei
M\"unchen, Germany}

\maketitle

\begin{abstract}
Observations by the MACHO collaboration indicate that a
significant fraction of the galactic halo dark matter may be in form of
compact objects with masses $M\sim 0.5M_{\odot}$. Identification of these
objects as red or white dwarfs is problematic due to stringent observational
upper limits on such dwarf populations.
Primordial black hole (PBH) formation from
pre-existing density fluctuations is facilitated during the cosmic 
QCD transition due to a significant decrease in pressure forces. 
For generic initial density perturbation spectra this implies that 
essentially all PBHs may form with masses close to the QCD-horizon scale,
$M_h^{\rm QCD}\sim 1M_{\odot}$.  It is possible that such QCD PBHs
contribute significantly to the closure density today. I discuss the status
of theoretical predictions for the properties of QCD PBH dark matter.
Observational signatures of and constraints on 
a cosmic solar mass PBH population are also discussed.
\end{abstract}

\begin{multicols}{2}
\section{PBH Formation during the QCD Epoch}

It is long known that only moderate deviations from homogeneity
in the early universe may lead to abundant production of PBH's 
from radiation \cite{Zel}. For a radiation equation of state (i.e. $p=\rho /3$, where
$p$ is pressure and $\rho$ is energy density) there is approximate
equality between the cosmic Jeans-, $M_J$, and horizon-, $M_h$, masses.
The ultimate fate of an initially super-horizon density fluctuation,
upon horizon crossing,
is therefore determined by a competition between dispersing
pressure forces and the fluctuation's self-gravity.
For fluctuation overdensities exceeding a
critical threshold at horizon crossing $(\delta\rho /\rho)_{hc} \geq 
\delta_c^{\rm RD}\approx 0.7$ 
\cite{Nie1} formation of a PBH with mass $M_{pbh}\sim M_h$ results.

The universe must have passed through a color-confinement quantum 
chromodynamics (QCD) transition at cosmic temperature $T\approx 100$MeV.
Recent lattice gauge simulations indicate that the transition between a
high-temperature quark-gluon phase and a low-temperature hadron phase
may be of first order \cite{Kan}, even though such simulations are still plagued
by limited resolution and problems to 
account for finite strange quark mass. A first order 
phase transition is characterized
by coexistence of high- and low-temperature phase at coexistence temperature
$T_c$. Both phases may exist in pressure equilibrium, $p_c^{qg}=p_c^h$ but with
different and constant (at $T_c$) energy densities, $\rho_c^{qg}-\rho_c^h=L$,
where $L$ is the latent heat. 
During phase coexistence adiabatic expansion of the universe
causes a continuous growth of the
volume fraction occupied by hadron phase
$(1-f_{qg})$, on the expense of quark-gluon
phase,  such that through the release of latent heat the universe is kept at $T_c$. 
The transition is completed when all space is occupied by hadron phase.

Consider a volume element of mixed quark-gluon- and hadron phase-
during phase coexistence. Provided a typical length scale of the volume
element is much larger than the mean separation between quark-gluon
and hadron phase (i.e. the mean hadron- or quark-gluon- bubble separation,
$l_s$) one may regard the volume element as approximately homogeneous. 
The average energy density of the volume element is
$\langle\rho\rangle =\rho_c^h
+f_{qg}L$ and continuously varies with the change of $f_{qg}$, whereas
pressure remains constant, $p=p_c^h$. Upon adiabatic compression of
mixed quark-gluon/hadron phase there is therefore no pressure
response, $v_s^{eff}=\sqrt{(\partial p/\partial\langle\rho\rangle)_s}=0$
\cite{Jed1}. Of course, 
the pressure response may only vanish if thermodynamic equilibrium 
is maintained. For rapid compression time scales or small 
compression amplitudes this may not be the case, 
whereas it is anticipated
that approximate thermodynamic equilibrium applies over a Hubble time
and order unity compression factors. During phase coexistence the universe
is effectively unstable to gravitational collapse for all scales exceeding $l_s$.
Note that a vanishing of $v_s^{eff}$, which was independently discovered
by \cite{Schmid}, may also have interesting non-gravitational 
effects on density perturbations.

These considerations have led me \cite{Jed1} to propose PBHs formed during the
QCD epoch 
from pre-existing initially superhorizon density fluctuations, such as leftover
from an early inflationary period of the universe,
as a candidate for non-baryonic dark matter. Fluctuations
crossing into the horizon during the QCD epoch experience a 
significant reduction of pressure forces over 
that regime of the fluctuation which exists in mixed phase. Since the PBH
formation process is a competition between self-gravity and pressure forces,
and $v_s=1/\sqrt{3}$ is constant during most other radiation dominated epochs,
the threshold for PBH formation should be smaller during the QCD epoch than
during other early eras, $\delta_c^{\rm QCD}<\delta_c^{\rm RD}$.
Only a slight favor for PBH formation during the QCD epoch may effectively
lead to the production of PBH on only approximately the QCD 
horizon mass scale $M_h^{\rm
QCD}\approx 2M_{\odot}(T_c/{\rm 100 MeV})^{-2}$. This holds true
for strongly declining
probability distribution functions for the pre-existing fluctuation overdensities.
For example, assuming Gaussian statistics, PBH formation is
dominated for $\delta\rho /\rho$ in the range $\delta_c$ and  
$\delta_c+ \sigma^2 /\delta_c$, where $\sigma$ is the 
variance of the Gaussian distribution. This range is very small,
$\sigma^2 /\delta_c \simle 10^{-2}$, if 
PBH mass density is not to exceed the present closure
density, $\Omega_{pbh}\simle 1$. For $\Omega_{pbh}\approx 1$
PBH formation during the QCD epoch is also a very rare event with only
a fraction $\sim 10^{-8}$ of horizon volumes collapsing to black holes.

The possible production of PBH during the QCD epoch is not a
completely new suggestion. In fact, in the mid seventies it was believed
that a QCD era was characterized by an ever-increasing production of
massive hadronic resonances. Such a \lq\lq\ soft \rq\rq 
(i.e. almost pressure-less) Hagedorn era
was argued to be suspect since overproduction of primordial
black holes seemed likely \cite{Chap}. In the eighties it was argued that the
long-range color force could lead to the generation of subhorizon density
fluctuations which in turn could collapse to 
planetary sized PBHs \cite{Craw}. Nevertheless, the simple properties
of mixed phase during a cosmic first-order transition and their possible
implications for PBH formation on the QCD horizon mass
scale have so far been overlooked.

Currently there are two groups attempting to simulate the PBH formation
process during a QCD transition with aid of a general-relativistic
hydrodynamics code \cite{Wil,Jed2}. Preliminary results by \cite{Jed2} 
verify the reduction
of PBH formation threshold during the QCD epoch.
Assuming a bag equation of state and phase
transition parameter, $L/\rho_c^{h}=2$, we have found a 
PBH formation threshold reduction, 
$\delta_c^{\rm QCD}/ \delta_c^{\rm RD}\approx 0.77$, for fluctuations entering
the horizon approximately during the middle of the phase transition.
Note that a canonical bag model
with total statistical weights of $g_{qg}=51.25$ and $g_h=17.25$ for
quark-gluon- and hadron- phases \cite{Ful}, respectively, predicts even larger
$L/\rho_c^{h}=2.63$, 
whereas lattice simulations may favor smaller $L$ \cite{DeTar}.
For fluctuations entering the horizon during the QCD epoch one typically
finds the evolution of the fluctuation into two different spatial regimes. 
An inner part of the fluctuation exists in pure guark-gluon phase 
$\rho >\rho_c^{qg}$ surrounded by an outer part existing in pure hadron-phase
$\rho <\rho_c^{h}$. The enhanced
density in the inner part of the fluctuation assists the collapse to a PBH.
This is in contrast to PBH formation during simple radiation dominated
eras, where the fluctuation's density distribution is continuous.
It has also been attempted to derive approximate analytic estimates
of the threshold reduction and PBH masses 
for PBH formation during the QCD epoch \cite{Card}. The model predicts that
$\delta_c^{\rm QCD}$ is minimized for fluctuations entering the horizon well
before the transition resulting into PBH masses considerably smaller than the
QCD horizon. 

\section{Theoretical Predictions for QCD PBH Scenarios}

It is valuable to advance the initial suggestion of possible
abundant production of PBH during the QCD epoch to a complete and
predictive scenario. 
I outline here to which degree this may be accomplished and briefly
describe the theoretical issues in QCD PBH scenarios.

{\bf Threshold reduction:}
The bias for forming PBHs almost exclusively on the QCD scale
is dependent on the PBH formation threshold reduction,
$\delta_c^{\rm QCD}<\delta_c^{\rm RD}$. Within the context of a bag model
equation of state for a first order transition 
preliminary results of numerical simulations 
confirm the proposed threshold 
reduction. For higher order QCD transitions threshold reduction could still
occur but would have to be verified by using accurate $\rho (T)$, $p(T)$, and
$v_s(T)$ determined from lattice gauge simulations.
Due to the duration of the QCD transition, $L/\rho_c^h\sim 1$,
threshold reduction will be of order unity. A very accurate
determination of $\delta_c^{\rm QCD}$ is only necessary if PBH formation
is efficient over a range in $\delta\rho /\rho$ which is also of order unity.
This is not the case for Gaussian statistics of the pre-existing density
fluctuations but may apply for non-Gaussian statistics.

{\bf Mass function:}
A crucial prediction of a QCD PBH dark matter scenario is the
average QCD PBH mass.
There is seemingly rough agreement of the QCD horizon mass scale
$M_h^{\rm QCD}\approx 2M_{\odot}(T_c/{\rm 100 MeV})^{-2}$ and the
inferred masses of compact objects in the galactic halo by the MACHO
collaboration, $M\sim 0.5M_{\odot}$. Nevertheless, currently there are large
uncertainties in the prediction for average QCD PBH mass, 
$\langle M_{\rm pbh}^{\rm QCD} \rangle$. 
Even incorrectly assuming $M_h^{\rm QCD}=\langle M_{\rm pbh}^{\rm QCD} 
\rangle$,
there is a factor eight uncertainty in $M_{\rm pbh}^{\rm QCD}$ depending on if
the horizon length is taken as
radius or diameter of a spherical horizon volume. 
The QCD equation of state, order
of the transition, and transition temperature are as yet not precisely
determined. The transition temperature may fall somewhere in the
range $200\, {\rm MeV}\simge T_c \simge 50\, {\rm MeV}$ implying a factor
sixteen uncertainty in $M_h^{\rm QCD}$, and probably equal uncertainty
in $\langle M_{\rm pbh}^{\rm QCD} \rangle$. Assuming a first order transition,
$\langle M_{\rm pbh}^{\rm QCD} \rangle$ may also depend on $L$ and the
equation of states above, and below, the transition point.
An accurate determination of $\langle M_{\rm pbh}^{\rm QCD} \rangle$
requires detailed and reliable lattice gauge simulation data. Approximate
trends may be obtained by using a bag equation of state.
A PBH mass function, as well as $\langle M_{\rm pbh}^{\rm QCD} \rangle$,
is obtained by convolving the distribution function for density contrast 
of the pre-existing density perturbations,
$\delta\rho /\rho$, with a scaling relation
associating final PBH mass with density contrast \cite{Nie2}.
The average PBH mass is thus also dependent on the statistics of the
density perturbations. Further, it has been shown that resulting PBH masses
are dependent on the fluctuation shape \cite{Bick}. These uncertainties are
particularly difficult to remove since they require knowledge about the
underlying physics creating density perturbations, presumably occurring
at a scale not accessible to particle accelerators.

{\bf Contribution to $\bf \Omega_{pbh}$:}
The contribution of QCD PBHs to the closure density at the present epoch
is dependent on the fraction of space which is overdense by more than
$\delta_c^{\rm QCD}$. COBE normalized, exactly scale-invariant $(n=1)$
Gaussian power spectra, imply negligible PBH production.
Gaussian blue spectra with $ 1.37\leq n \leq 1.42$ predict $\Omega_{pbh}$ 
in QCD PBHs in the range $10^{-5}$ to $10^3$ \cite{Card}. 
Such spectral indices are
consistent with cosmic microwave background observations
\cite{Benn}.
Nevertheless, blue spectra resulting from inflationary epochs have been
shown to generically be non-Gaussian, skew-negative \cite{Bull}. 
Density perturbations with an exactly scale-invariant,
COBE normalized power spectrum,
but with a non-Gaussian, skew-positive distribution tail, may yield
$\Omega_{pbh}\sim 1$. One argument against QCD PBH dark matter
is the degree of fine-tuning involved for obtaining $\Omega_{pbh}\sim 1$.

{\bf Accretion around recombination:}
It is long known that black holes may efficiently accrete
after the epoch of recombination \cite{Carr}. 
Whereas accretion does not appreciably
change the black hole masses, conversion of accreted baryon rest mass
energy into radiation may produce substantial radiation backgrounds.
The presently observed X-ray and/or UV
backgrounds may be incompatible with a population of PBHs with mass
$M_{pbh}>10^4 M_{\odot}$ and $\Omega_{pbh}>0.1$ \cite{Carr}. A population of
$M_{pbh}\sim 1 M_{\odot}$ PBHs with large $\Omega_{pbh}$ is consistent
with the observed X-ray and/or UV backgrounds. Accretion of baryons
on PBH shortly before the epoch of recombination
may produce distortions in the blackbody of the cosmic microwave background
radiation. PBHs with $M_{pbh}\sim 1 M_{\odot}$ would accrete at the Bondi
rate, with Thomson drag inefficient. Tidal interactions between the 
accreting gas and
neighboring PBHs would lead to the transfer of angular momentum
and the formation of disks around the PBH. Preliminary results of an investigation
of PBH accretion before recombination
indicate that the resulting blackbody distortions would be below the
current FIRAS limit.

{\bf PBH formation during other epochs:}
Efficient PBH formation during the QCD era may, in principle, imply
formation of PBHs during other epochs as well. For example, 
during the $e^+e^-$-annihilation there is a decrease in the speed of sound
which may result in a bias to form PBHs on the approximate horizon scale
of this era. Further, for power spectra of the underlying density distribution
characterized by $n>1$ QCD PBH formation may be accompanied by PBH
formation at earlier times on mass scales $M \ll M_h^{\rm QCD}$.
It is important to verify that such PBHs do not violate observational 
constraints \cite{Green}.

\section{Observational Signatures of QCD PBH Dark Matter}

Ultimately, only by observational technique the existence of a 
population of QCD PBH may be established. It is therefore important to establish
the observational signatures of QCD PBH dark matter. Particular emphasis
is laid on observations which may be performed in the not-to-distant future.

{\bf Galactic halo microlensing searches:}
The recent results of microlensing searches for compact, galactic
halo dark matter by the MACHO collaboration 
\cite{Alco} provide some motivation
for QCD PBH dark matter. Low event statistics as well as uncertainties about
the halo model which is to be adopted result in fairly large ranges for
the average MACHO mass, $0.1M_{\odot}\simle M\simle 1M_{\odot}$, 
and halo dark matter fraction
provided by MACHOs, $f_M\simge 0.2$. 
The error bars may be
reduced by increasing the number of observed microlensing events and
observing towards several line-of-sights (e.g. towards the Large and Small
Magellanic Clouds). Nevertheless, it will not be possible by only the 
observational MACHO project to determine an accurate mass function.
Only in combination with follow-up observations, such as by a space
interferometry sattelite, 
degeneracy between MACHO lens mass, distance,
and projected velocity may be lifted and
a mass function may be determined.

{\bf Alternative interpretations of the MACHO results:}
The inferred masses of MACHOs are close to those of stars, stellar
remnants, or brown dwarfs. The most straightforward interpretation
of the observations by the MACHO collaboration are that baryonic objects
have been detected. 
However, one has to resort to fairly extreme galactic models
in order for a characteristic MACHO mass of $M\simle 0.1M_{\odot}$ 
to be consistent with the observations and for brown dwarfs to remain
a viable interpretation for the lenses.
A significant contribution to the halo dark matter by
red dwarfs seems ruled out by observations of the Hubble deep field \cite{Flynn}.
Halo white dwarfs with halo dark matter fractions exceeding $f_M\simge 0.1$
seem also in conflict with observations of the Hubble deep field, even though
this constraint is dependent on somewhat uncertain
white dwarf ages and cooling curves \cite{Graff}.
In addition, it has been argued that the light
which would be
emitted by the progenitors of abundant halo white dwarf populations 
has not been observed in deep galaxy surveys \cite{Charl}. 
It has been suggested that the lenses responsible for the observed microlensing
are not within the halo, but within a warped or thick galactic disk. Such
scenarios may possibly be rejected by microlensing observations on more than
one line of sight. 
There are other more, or less, radical interpretations
of the results of the MACHO collaboration. It is important, 
not only for the viability of QCD PBH dark matter, to establish, or rule out,
these alternative interpretations.  

{\bf Quasar microlensing:}
The optical depth for microlensing of distant quasars by a cosmic
component of compact, solar mass objects 
with $\Omega_{c}\sim 1$ is remarkably large.
In fact, a constraint of $\Omega_{c}\simle 0.2$ for 
a population of compact
objects with masses $M_c\sim 1M_{\odot}$ has been derived from observations
of broad line radiation- to continuum radiation- flux ratios of $\sim 100$ quasars
\cite{Dal}.
This limit relies on the assumption that most continuum radiation is emitted
from within a compact $\simle 0.1$pc region in the center of the quasar, whereas 
the broad line radiation emerges from a much more extended region around the
quasar. The limit is independent of the clustering properties of the compact
objects. There is as yet no conclusive model for quasar variability. It has thus
been proposed that quasar variability is due to microlensing of an $\Omega_c\sim
1$ component of compact objects with $M_c\sim 10^{-3}M_{\odot}$ \cite{Haw}.
QCD PBH dark matter may therefore be constrained by large, homogeneous
samples of quasar observations, such as expected to result from the Sloan
Digital Sky Survey, hopefully accompanied by an improved understanding of the
physics of quasars.

{\bf Gravitational wave detection from PBH binaries:}
It has been shown that a fraction $10^{-2}-10^{-1}$ of QCD PBHs may
form in PBH binaries \cite{Naka}. This values is in rough
agreement with the fraction of binaries 
observed by the MACHO collaboration.
Gravitational waves emitted during PBH-PBH mergers are above
the expected detection threshold for the LIGO/VIRGO interferometers
when occurring within a distance of $\sim 15$Mpc.
For galactic halos consisting exclusively of QCD PBH dark matter with
$M_{pbh}\sim 0.5M_{\odot}$ this implies that up to a few mergers
per year may be detected by the 
next generation gravitational wave interferometers \cite{Naka}.
It is particularly encouraging that the gravitational wave signal is sensitive
to the masses of PBH within the binary. One may hopefully also 
distinguish between
neutron star and black hole binaries. Establishing the existence of
black holes with masses well below the upper mass limit for neutron stars
may strongly argue in favor of primordial black holes.
 
{\bf Galactic disk accretion:}
Limits may be placed on galactic halo PBH number densities by the
accretion induced radiation which may be observed when a halo PBH passes
through the galactic disk in the solar vicinity \cite{Heck}. Nevertheless, even the
$\sim 10^8$ objects which will be observed within the Sloan Digital Sky Survey
will not provide sufficient statistics to establish, or rule out, an all QCD PBH
halo with masses as small as $\sim 1M_{\odot}$.

\section{Conclusion}

QCD PBHs may be an attractive dark matter candidate. I have outlined here
to which degree accurate predictions for the properties of QCD PBH
dark matter may be made. Most uncertain is the contribution to $\Omega$
of such objects since it relies on knowledge about the underlying density
perturbations on mass scales not accessible to cosmic microwave background
radiation observations. Predicting QCD PBH mass functions beyond the
approximate equality between MACHO masses and the QCD horizon mass
may improve with detailed numerical simulations of the PBH formation process
and future results of lattice gauge simulations for the QCD equation of state.
A combination of observational techniques, such as galactic microlensing
searches, quasar microlensing searches, and gravitational wave interferometry
may point towards the abundant existence of such objects. Ultimately,
the unambiguous detection of a black hole well below the
maximum mass for neutron stars may argue strongly for its primordial nature.

\section*{Acknowledgment}   

I am very grateful to my collaborator J.C. Niemeyer, for his permission to
present preliminary results of our work in these proceedings, and for many 
useful discussions. I am also indebted to T. Abel, C.R. Alcock, G.M.
Fuller, and S.D.M. White for their encouragement.

\end{multicols}

\end{document}